\documentclass[a4paper,11pt,colorlinks=true,urlcolor=blue,anchorcolor=blue,citecolor=blue,filecolor=blue,linkcolor=blue,menucolor=blue,linktocpage=true,pdfproducer=medialab,pdfa=true]{article}
\pdfoutput=1 
\usepackage{orcidlink} 

\usepackage{jcappub} 
\usepackage[T1]{fontenc} 
\usepackage{xcolor}
\usepackage{soul}   
\usepackage{comment}
\usepackage{subcaption}
\usepackage{dsfont}
\usepackage{physics}
\usepackage{amssymb}
\usepackage[normalem]{ulem}

\title{Influence of photometric galaxies redshift distribution in BAO estimation}

\author[a]{Paula S. Ferreira\orcidlink{0000-0002-7540-040X},}
\author[a,b]{Ribamar R. R. Reis\orcidlink{0000-0003-1339-2106}}
\affiliation[a]{Instituto de Física, Universidade Federal do Rio de Janeiro,\\
Av. Athos da Silveira Ramos, 149 - Cidade Universitária, Rio de Janeiro, Brasil}

\affiliation[b]{Observatório do Valongo, Universidade Federal do Rio de Janeiro\\
Ladeira Pedro Antônio, 43, Centro,  Rio de Janeiro,
Brasil}
\emailAdd{psfer@pos.if.ufrj.br, ribamar@if.ufrj.br}

\abstract{Photometric surveys are good solutions for large-scale structure studies. The Baryon Acoustic Oscillations (BAO) benefit from photometric redshift survey observations due to faster coverage and a higher number of observed objects. In the present study, we use the Dark Energy Survey Year 3 catalog of Luminous Red Galaxies (LRG) to incorporate the realistic galaxies' redshift Probability Distribution Function (PDF) into the correlation function cosmological model. We used four different photo-z estimators \texttt{ANNz2}, \texttt{BPZ}, \texttt{ENF}, and \texttt{DNF} to compare how they affect the BAO feature constraint, the catalogs called \texttt{Pz Cats}.  Moreover, each algorithm included two sample selections based on distinct PDF shapes; one where the PDFs are nearly Gaussian and another opting for the least noisy PDFs with a pronounced peak. Our investigation identifies whether our samples detect BAO using the correlation function's polynomial parametrization. 
Following this parametrised model, we estimated the shift parameter $\alpha$ for the \texttt{ANNz2} three cuts and the \texttt{DNF} full samples. We found that the BAO from \texttt{ANNz2} Gaussian sample selection is closer to the fiducial \texttt{Planck} 18 $\Lambda$CDM model.
Later, we computed the correlation function $\xi_\perp(z_p)$ by getting the bin pairs transversal to each other using \texttt{CAMB}. The kernel window function is the $f(z|z_p)$ which is the selection of the PDF value when the photometric redshift is nearly the same as the spectroscopic redshift estimated by the matched spectroscopic sample. For compatible $z_{\rm eff}$, we concluded that the shape of the galaxy redshift PDF could shift the BAO feature position either by including the PDF in the model or not. We also learnt that, given the same spectroscopic sample, \texttt{ANNz2} estimator with its respective selection samples outperforms other estimators for most parameters examined. When the dark energy equation of state parameter, $w_0$, is considered, \texttt{DNF} emerges as the optimal algorithm, provided it has sufficient statistical data. Our analysis recommends that upcoming photo-z survey collaborations incorporate multiple photo-z estimation algorithms in their cosmological inference process; this approach will facilitate comprehension of systematic effects on various parameters.}
\begin{document}
\maketitle
\flushbottom

\keywords{Cosmology \and Large Scale Structure \and Baryon Acoustic Oscillations \and Photometric surveys}

\section{Introduction}

The Baryon Acoustic Oscillations (BAO) were spherical sound waves formed by the oscillations in a baryon-photon gas that resulted from density fluctuations seeded by quantum fluctuations amplified by inflation \cite{peebles1980large,hu2002cosmic}. The oscillations are described as a competition between pressure and gravitational collapse. As the universe expands and cools down, charged particles reach the ideal temperature to form neutral atoms. This causes photons not to scatter anymore and allows them to travel freely throughout the universe. Consequently, the last sound wavefront from BAO is frozen as a spherical pattern at the sound horizon of such decoupling, around $r_d:=r_s(z_d)=150$ Mpc for the $\Lambda$CDM model (the standard cosmological model) for a flat scenario where dark energy dominates constituting nearly 70\% of the universe, followed by cold dark matter with 30\% dominance. These regions with higher density are expected to be slightly preferred for galaxy formation, inducing a higher correlation between galaxies separated by this distance.

The BAO feature can be constrained by galaxy surveys or Cosmic Microwave Background (CMB) observations. The main type of data set comes from galaxy surveys that can be spectroscopic or photometric. Spectroscopic surveys are precise since they rely on spectrographs. For each object, the light goes through optical fibres or slits onto the spectrograph. In some surveys, hundreds or even thousands of fibres can be placed in the telescope focal plane, allowing for simultaneous observation of multiple galaxies. This process of obtaining the spectra for distant galaxies is slow, e.g., it used to take around one hour per galaxy for the Sloan Digital Sky Survey (SDSS) \cite{Shu_2012} while observing Luminous Red Galaxies (LRG). The Dark Energy Spectroscopic Instrument (DESI), the SDSS's successor, is capable of observing more galaxies because of its robotic optical fibre \cite{silber2022robotic}. However, despite the great improvement in spectroscopic observations, the new instrument is also affected by fibre collisions that suppress power suppression in angular pair counts, but possibly mitigated by a collision weights \cite{bianchi2024characterization}.
These issues complicate observations of many targets, which is vital for cosmological studies.

The other survey strategy is a photometric survey. The instruments now collect information about an object from a set of filters covering some wavelength range. Photometric surveys can cover a wide area of the sky observing hundreds of thousands of objects simultaneously, including fainter objects, different from the optical fibre for each object like the spectrographs. This is advantageous for cosmology, but losing the resolution from spectroscopy. The filters are usually broad, and the instruments are built with a limited number of filters. SDSS for instance has filters $u,g,r,i,z$ \cite{fukugita96}. The filters $g,r,i,z$ were later used by the Dark Energy Survey (DES) \cite{drlica2018dark} plus the $Y$ filter, which covers the visible and part of the near-infrared light together. Few wide area surveys are conducted using a higher number of narrow band filters to circumvent this particular issue, such as  Javalambre-Physics of the Accelerated universe Astrophysical Survey (J-PAS) \cite{Benitez:2008ts,benitez2014}, Southern Photometric Local Universe Survey (S-PLUS) \cite{MendesdeOliveira:2019hfl}, Legacy Survey of Space and Time (LSST) \cite{collaboration2012large}, the \textit{Euclid} Wide Survey \cite{2025A&A...697A...1E}, the \textit{Roman Space Telescope} \cite{eifler2021cosmology} and \textit{China’s Space Survey Telescope} (CSST) survey \cite{zhan2021wide,miao2024forecasting}.

The redshift of a spectroscopic survey is measured directly from the galaxy spectrum compared to some known spectra measured in a laboratory. Photometric redshift is computed by estimating the best redshift based on a sample of known spectroscopy and photometry, the variables used to compare the samples are usually magnitudes. The methods are either template fitting (when the photometric redshift (photo-z) estimator has a group of known objects) and/or empirical and theoretical models (which will always be the reference to any estimation or a trained-based method, where some objects have known spectroscopic redshift and are the training sample).  The precision of a photometric redshift (photo-z) increases when there is enough spectroscopic reference to some of the objects observed or having as many filters as possible to be close to a spectroscopic observation. 

The challenge of finding the BAO feature using photometric surveys lies in the fact that the redshift of each object has a significant uncertainty. The angular BAO scale is less affected by redshift uncertainty, so most of the constrained BAO feature by photo-z surveys are transverse to the line of sight of the spherical signal.  The two ways of measuring the transverse signal are using the angular power spectrum or the angular correlation function \cite{peebles1973statistical}. This was later used by \cite{huterer2001angular} with the Edinburgh/Durham Southern Galaxy Catalogue Cl (EDSGC). Using the SDSS photo-z sample \cite{seo2012acoustic} measured the BAO using the angular power spectrum. After them, the DES Collaboration published their results within six years of data collection \cite{abbott2024dark}. 

In order to find the BAO feature in 3D from photo-z\cite{dodelson2002three}  used the angular power spectrum $C_\ell$ to get the 3D power spectrum $P(k)$, but the authors could not find the BAO signal. The usual assumption is that the photo-z errors are the standard deviation of a Gaussian distribution. This information is included in the number density as a function of redshift \cite{huterer2001angular}, \cite{ross2017optimized}, \cite{chan2018bao}. However, the redshift probability distribution function (PDF) obtained from the photo-z estimator is not a perfect Gaussian distribution, some can even be multi-modal. A more accurate method should include the information of the real PDFs obtained. \cite{asorey2016galaxy} tested with simulations the approach with the real PDFs for Large Scale Structure (LSS) analysis without the intention of finding the BAO. They found an increase of 1.67 times in precision compared to assuming some statistics to the PDFs. 

Later, \cite{chan2022clustering} proposed including the PDFs information and crossing photo-z bins in order to get the BAO signal. \cite{chan2022dark} used the DES Year 3 data set to apply \cite{chan2022clustering} method. They gave evidence of the projected correlation function $\xi_p(s_\perp)$ is sensitive to realistic PDFs. However, we still need to compare the performance of different PDFs estimated by different algorithms. 

In this work, we are interested in comparing the performance of photo-z estimators in measuring the BAO from their redshift PDFs outputs. We propose an approximate estimation of $\xi_p$ suggested by \cite{chan2022clustering}, but using the whole spectroscopic sample we used to train the photoz estimators. Furthermore, we also try to understand the implications of forcing a selection cut based on the galaxies' PDF shape, for that we try to get the smoother PDFs available in the catalogs (\texttt{Pz Cats}) we constructed and are described in the companion paper \cite{ferreira2025}.

The structure of the present study starts with the explained methods, including how we obtained the random catalog in section \ref{sec:methods}. Next, we give the detailed selection cuts in section \ref{sec:cuts} from \cite{ferreira2025}. After that, we explain the model description to get $\xi(s_\perp)$ in section \ref{sec:cosmology_model}. In section \ref{sec:results}, we show our results, and finally, the conclusion is in section \ref{sec:conclusion}.

\section{Methods}\label{sec:methods}

\subsection{Data Set}

Artificial Neural Networks (ANNs), such as \texttt{ANNz} \cite{collister2004annz}, are training-based methods used to estimate photometric redshifts ($z_p$) by mapping input variables (e.g., magnitudes $\mathbf{m}_k$) to outputs ($z_p$ and PDFs) through neurons and minimizing a cost function 
$E = \sum_k \left[z_{p}(\mathbf{w},\mathbf{m}_k)-z_k\right]$.
Improved versions like \texttt{ANNz2} \cite{sadeh2016annz2} employ randomised regression with machine learning methods (MLMs) to enhance $z_p$ PDF estimation. Additionally, nearest neighbour methods, including Euclidean Neighbourhood Fitting (ENF) and Directional Neighbourhood Fitting (DNF) \cite{de2016dnf}, estimate redshifts by considering distances $D = \sqrt{\sum_i^n (m_i^t-m_i^p)^2}$, \texttt{DNF} looks at the nearest neighbour in the magnitude space separated by $DN = D^2\sin ^2 \alpha$, where $\alpha$ is the angle separation between two magnitudes. Bayesian methods like \texttt{BPZ} \cite{benitez2000bayesian} use template fitting to estimate $p(z|\mathbf{m}_0)$, leveraging Bayesian probability theory. These techniques, applied with training, testing, and evaluation sets, ensure accurate redshift estimation and PDF construction, the detailed description of the catalogs generated can be found in \cite{ferreira2025}.

We got the LSS data set from the DES Collaboration in their Y3 analysis (DESY3, from now on) BAO sample \cite{abbott2022dark}.  In terms of systematic effects, we are not changing the sample itself, just using the re-estimated photo-z from \cite{ferreira2025} using the magnitude information measured by the Dark Energy Camera (DECam). The selection and imaging systematics are the same as in \cite{10.1093/mnras/stad2402}. We estimated the photo-z using the catalogs from \texttt{Pz Cats}, a set of catalogs described in detail in \cite{ferreira2025}: 
 \texttt{BPZ} \cite{benitez2000bayesian}, \texttt{ANNz2} \cite{sadeh2016annz2}, \texttt{ENF/DNF}\cite{de2016dnf}, the files are available at \hyperlink{https://doi.org/10.5281/zenodo.15191620}{https://doi.org/10.5281/zenodo.15191620} in the ZENODO \cite{zenodo} open data repository. Our results showed that a selection of galaxies based on their PDFs is a robust choice compared to the spectroscopic reference sample. The effective redshift of the samples are compatible with each other and with the DES sample, the difference is that the photometric sample redshift distribution alters depending on the cut and as a consequence their uncertainty. In section \ref{sec:cuts}, there is the detailed information of each sub-samples, where the effective redshift $z_{\rm eff}$ is described by \cite{ross2017optimized}. Further details can be found at \cite{ferreira2025}.

\subsection{Random catalog}
For clustering cosmology, we require a random set of galaxies. The random catalog must be distributed in the sky in a random distribution. For that, the random catalog is based on the survey's footprint. 

Because we used a different training set than the DESY3 Collaboration, the photo-z distribution of the random sample is not the same as the estimated from \texttt{Pz Cats}. The best solution was to randomly construct the redshift distribution for the random catalog that follows the resulting \texttt{ANNz2}, \texttt{BPZ}, and \texttt{ENF/DNF} distributions. For that, we used DES \cite{abbott2022dark} random catalog and selected the redshift using the Metropolis sampling method: for an initial redshift guess, the algorithm rejects or accepts the value that fits the expected distribution function. 

The selection described does not impact the core analysis because the fundamental purpose of the random catalog is not to replicate the exact, clustered redshift distribution of the data, but to represent the underlying, smooth selection function of the survey against which clustering is measured. As established in the literature, the random catalog must trace the survey's angular and radial window without any intrinsic clustering \cite{Anderson_2012, Padmanabhan_2007}. Its primary role is to characterize the survey geometry and selection probability, requiring a large number of points uniformly sampled within the defined footprint \cite{Wang_2013, abbott2022dark, Ferrero_2021}. Therefore, while the angular mask is fixed, the specific method for assigning redshifts to the randoms — as long as it generates a smooth distribution that samples the same volume —is a flexible aspect of the pipeline and does not bias the clustering measurement. 

\subsection{Mocks}
We used \texttt{CoLoRe} \cite{ramirez2022colore} to construct the mocks to validate our results.  The mocks are the realisations of the given survey if the observed redshift matched the spectroscopic redshift sample of reference. We chose the Log-Normal mock type because it is less time-consuming and applicable to $z>0.4$ without problems of non-linearities. The DES Collaboration used log-normal mocks from \texttt{FLASK} \cite{xavier2016flask} for model-fitting when using $C_\ell$s, because their N-body simulation mocks from \texttt{MICE} \cite{ferrero2021dark} had replicated structure problems, which is explicitly written in \cite{abbott2022dark}. The cosmological model for the mocks' construction is based on Planck 18 \cite{collaboration2020planck} results and \texttt{ANNz2}'s $n_{DES}(z_s)$ (described in \ref{sec:ndes}). The simulation has a 1024 grid size and 72 million galaxies for the whole sky. The redshift distribution is based on the re-sampled distribution from Eq.~(\ref{eq:n_spec_des}). We used the mask \cite{rosell2022dark} available at \href{http://desdr-server.ncsa.illinois.edu/despublic/y3a2_files/baosample/DESY3_LSSBAO_MASK_HPIX4096NEST.fits}{DES Data Management}. 

\begin{table}
    \centering
    \begin{tabular}{c}
        Mock realization parameters\\
        \hline
         $z_{min}=0.1$\\
         $z_{max}=2.0$\\
         $N_{grid}=1024$\\
         $N_{side} = 64$\\
         Angular resolution = $0.92^o$\\
         Tracer Kernel: $\Theta(z<0.1)$\\
         \hline
    \end{tabular}
    \caption{\texttt{CoLoRe} parameters used for mock ralizations..}
    \label{tab:mocks_params}
\end{table}

Lastly, the bias function is based on a best fit result using the DES Science Verification (SV) data by
\cite{salvador2019measuring} and \cite{crocce2016galaxy}\footnote{This bias function was used to determine the $b(z)$ in the $z_{\rm eff}$ equation.}:
\begin{equation}\label{eq:biasf}
    b_{best}(z) = 0.98 + 1.24z-1.72z^2 + 1.28z^3.
\end{equation}
We compared the bias function from SV \cite{salvador2019measuring,crocce2016galaxy} \texttt{ANNz2} and our \texttt{ANNz2} output, the difference between the galaxy bias was minimum (see Appendix~\ref{ap:SV_bias}).
 In table \ref{tab:mocks_params}, we show other important \texttt{CoLoRe} parameters for our pipeline. $z_{min}$ and $z_{max}$ are the redshift minimum and maximum values, $N_{grid}$ is the number of grid cells the box is divided, $N_{side}$ which leads to the angular resolution of $0.92$, and the tracer kernel is a Heaviside function. The simulations were done in the full and then we cut the sample as the correspondent survey footprint. 

In the end, we got 100 mocks with $\sim 7.8 \times 10^6$ galaxies each. We combined the one random mock set available by the DES to construct the mocks random catalog using the same pipeline for the observed set (re-using the collaboration's random sky distribution and resampling to match the redshift distribution). 
\subsection{The expected redshift distribution \texorpdfstring{$n_{DES}(z_s)$}{nzdes}}\label{sec:ndes}
The spectroscopic redshift distribution of the whole sample is based on the galaxies used as a reference from the spectroscopic sample. The distribution is computed with the distribution of the spectroscopic set $n_{DES}(z_s)$, the photometric redshift of the spectroscopic sample $n(z_p)$ computed with the specific algorithm, the photo-z distribution of the whole survey $n_{DES}(z_p)$, and $g(z|z_p)$, the value of the PDF where spec-z is equal to the photo-z of the matched galaxies. This is the resampled distribution, written as:

\begin{equation}\label{eq:n_spec_des}
    n_{DES}(z_s) \approx \int \dd z_p \frac{n(z_s) n_{DES}(z_p)}{n(z_p)}g(z|z_p)
\end{equation}
\subsection{Multi-modal PDF criterion}\label{sec:cuts}

Some PDFs are multi-modals; they present multiple peaks. This is an indication of a bad photo-z estimation, which can degrade the BAO signal. Unfortunately, one cannot eliminate all the objects with secondary peaks, but it is possible to exclude the most noisy ones.

Here, we briefly describe the criteria we discussed in \cite{ferreira2025}. We computed the number of peaks per PDF and selected galaxies that contain one peak that cannot be larger than 30\% of the main peak, which is 1.55$\sigma$ from the mean if the distribution was considered Gaussian, this ensures we are not too conservative and keep realistic results. In other words, we avoid distributions with many high secondary peaks. Previously, we tested distributions without any secondary peaks, but this reduced sample size significantly depending on the photo-z estimator.

We can also assume that the PDFs are Gaussian. We selected PDFs close to a Gaussian distribution. We used the statistical moments $\mu_n$ using the distributions $PDF(z_p)$:
\begin{equation}
    \mu_n = \int\limits_{-\infty}^{\infty} z_p^n PDF(z_p) \dd z_p,
\end{equation}
where $n$ is the statistical moment, we choose to use the second moment to classify the distribution. For a Gaussian, the second moment is the sum of the average value squared and the variance ($\mu^2+\sigma^2$). The mode and the mean are the same in a normal distribution, so we use the $z_p$ output of the estimators as $\mu$ and its respective error as $\sigma$. Besides the previous specifications, it also obeys the Multi-modal criterion. 

The error estimation of the samples is described in \cite{ferreira2025}, table \ref{tab:cuts} shows each sub-sample and the full sample results. We will henceforth use the term "estimator" to describe the sample including all galaxies. When referring to "estimator $+$ Gaussian", this denotes the sample where the galaxies' PDFs are close to a Gaussian distribution. Lastly, the "estimator $+$ Small Peaks/Sp" represents the criterion for selecting PDFs with minimal noise. 
We summarise the sub-samples as follows:

\begin{itemize}
    \item Full sample: no restriction to the PDFs.
    \item Gaussian: selection of PDFs that are close to a Gaussian distribution.
    \item Small-peaks: accepts PDFs whose secondary peaks are not higher than the mode. 
\end{itemize}

\begin{table}[ht]
    \centering
    \begin{tabular}{c|c|c|c|c}
       Photo-z estimator & Method  & No. of galaxies & $z_{\rm eff}$ &$\sigma_z^{68}$\\
       \hline
       \texttt{ANNz2} & Full sample & $7,081,993$ &0.856 &$ ^{+0.007}_{-0.021}$\\
       \hline
        \texttt{ANNz2} & Gaussian PDFs & $2,931,677$& $0.817$& $^{+0.014}_{-0.031}$\\
        \hline
        \texttt{ANNz2} & Small peaks & $5,133,775$ & $0.799$ &$^{+0.009}_{-0.022}$\\
        \hline
        \texttt{ENF} & Full Sample & $7,081,993$ &0.854 &$^{+0.018}_{-0.016}$\\
        \hline
        \texttt{ENF} & Gaussian PDFs & $299,427$ & $0.834$ & $^{+0.040}_{-0.084}$\\
        \hline
        \texttt{ENF} & Small peaks & $2,256,616$ & $0.768$ & $^{+0.046}_{-0.055}$ \\
        \hline
        \texttt{DNF} & Full Sample & $7,081,993$ &0.819 & $^{+0.054}_{-0.016}$\\
        \hline
        \texttt{DNF} & Gaussian PDFs & $2,111,461$ &0.868 & $^{+0.116}_{-0.121}$\\
        \hline
        \texttt{DNF} & Small peaks &$2,456,190$ & $0.761$ & $^{+0.033}_{-0.142}$ \\
        \hline
        \texttt{BPZ} & Full Sample & $7,081,993$ &0.848 &$^{+0.009}_{-0.039}$\\
        \hline
        \texttt{BPZ} & Gaussian PDFs & $3,973,671$ & $0.812$ & $^{+0.016}_{-0.045}$\\
        \hline
        \texttt{BPZ} & Small peaks & $5,904,860$ & $0.811$ & $^{+0.008}_{-0.042}$
    \end{tabular}
    \caption{All the sample cuts for each estimator, their size, $z_{\rm eff}$, and $\sigma_{z}^{68}$ (the 68th percentile of the redshift uncertainty $\sigma_z = \frac{(z_p-z)}{(1+z)}$, where $z$ is the spectroscopic redshift).}
    \label{tab:cuts}
\end{table}

\section{Transverse BAO from \texttt{Pz Cats}}
After introducing the cuts, we analyse, for each criterion and photometric redshift estimator, the correlation function with respect to the distance perpendicular to the line-of-sight ($\xi(s_\perp)$). 
We will fit a parametrised model in order to see the shift to the BAO for each sample cut, the change in the correlation function gives motivation to go through a thorough analysis based on the PDFs influence the power spectrum transformation into the correlation function. The dependence to the amplitude, feature width and other configuration will not be taken into account.

Furthermore, the BAO feature does not precisely correspond to the sound horizon when extracted from the power spectrum or correlation function. This discrepancy arises because the oscillations or peaks present are not sufficiently sharp to align perfectly with an exact measurement of $r_{drag}$ \cite{sanchez2008best}. Nevertheless, for the sake of comparison, we include our result along the Planck 18's sound horizon \cite{collaboration2020planck} and DESY3's \cite{abbott2022dark} results.

Pair counting was conducted using \texttt{Corrfunc} as described in \cite{sinha2020}. The maximum allowable line-of-sight distance was set at $s_\parallel=120$ Mpc/h, with a fine separation of $\Delta s_\parallel=1$ Mpc/h. The range for the transverse separation was $20<s_\perp<175$ Mpc/h, with a broader separation of $\Delta s_\perp=5$ Mpc/h. This maximum line-of-sight distance does not significantly exceed the $\Lambda CDM$ BAO feature, primarily because our focus is not on detecting signals associated with it but rather on the transverse configuration. The finer separation $\Delta s_\parallel$ ensures precise integration, accommodating fluctuations arising from redshift uncertainties; this approach and its validity have been tested by \cite{chan2022clustering}. Conversely, the transverse distance utilizes a broader separation due to its reduced sensitivity to photometric redshift errors, a methodology also supported by the findings of \cite{chan2022clustering} and implemented in \cite{rosell2022dark}.

Here, we have fitted \cite{sanchez2011tracing} adapted to $\xi(s_\perp)$:
\begin{equation}\label{eq:sanchez}
    \xi(s_\perp) = A + B s_\perp^\gamma+C\,exp\left(\frac{-(s_\perp-s_\perp^{fit})^2}{2\sigma^2} \right),
\end{equation}
where $A$ is an amplitude parameter, $B$ and $\gamma$ define the function's decay, $C$ is the amplitude of the Gaussian peak where we expect to find the BAO feature, $\sigma$ the width of the BAO peak, and $s_\perp^{fit}$ the position of the feature.

\begin{figure}
    \centering
    \includegraphics[width=\linewidth]{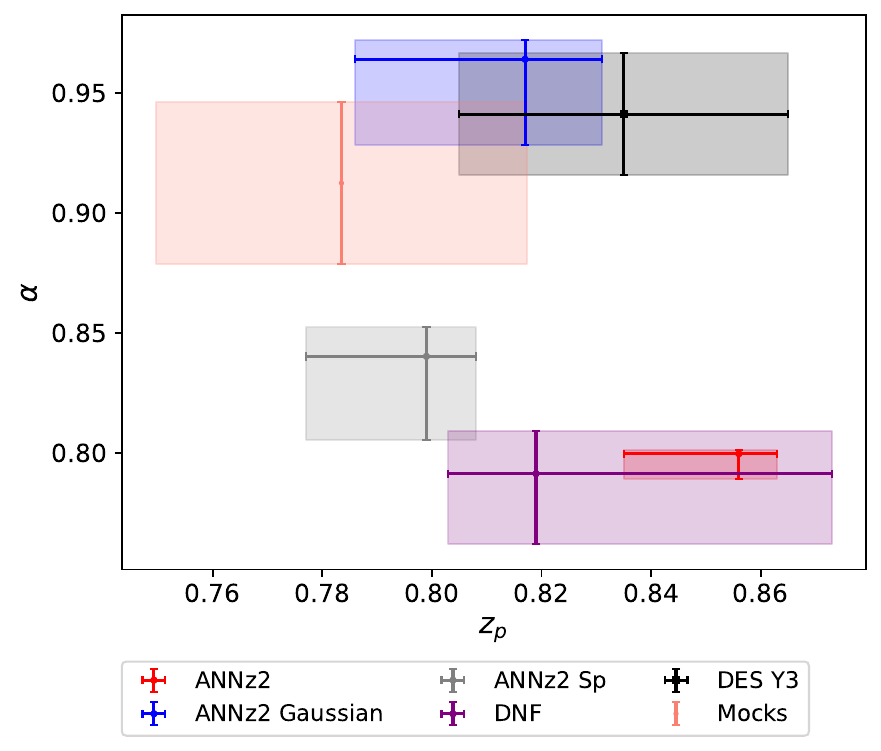}
    \caption{The shift parameter $\alpha$ is depicted for several methods: \texttt{ANNz2} in red, \texttt{ANNz2} with a Gaussian distribution in blue, and \texttt{ANNz2} Sp in gray. Additionally, \texttt{DNF} is shown in purple, the DES Y3 result as per \cite{abbott2018dark} is in black, and mock data is represented in salmon. Error bars are provided for both the shift parameter $\alpha$ and $z_p$ to illustrate the measurement uncertainties.}
    \label{fig:bao_post}
\end{figure}

Our main concern is: given the same statistical conditions, but different photo-z estimations, how does the BAO position shift? To answer the question, we can see in Fig.~\ref{fig:MCMC} the parameter space results for the four cases in which the BAO feature was detected, that is, $C>0$. Because the other parameters $\sigma$ and $s_\perp^{fit}$ impose a normal distribution, the BAO is only present if $C>0$, otherwise the peak would be a through.

The MCMC analysis to estimate the parameters were done using a bootstrap to construct a covariance matrix from 50 bins (using bootstrap) for each of the samples that had the BAO feature (see Appendix~\ref{ap:sanchez} for details). To obtain the sound horizon from the parametrised function from \cite{sanchez2011tracing}, we used the same correction from \cite{Ferreira:2023ecc}, which is a comparison between the correlation function value at $s^{fit}_\perp$ from the best-fit and each one of the 50 bins.  This can be written as:
\begin{equation}
    \tilde{\alpha}^i = \frac{\xi(s_\perp^{fit})-\xi^{i}(s_\perp^{fit})}{\int \sqrt{|\xi(s_\perp^{fit})-\xi^{i}(s_\perp^{fit})|^2} \rm{d}s_\perp}
\end{equation}
 and the final BAO position is:
\begin{equation}
     s_{\perp}^{\rm BAO}=(1+\mathrm{Mode}(\tilde{\alpha}^i))s_\perp^{fit}.
 \end{equation}
This projection effect correction is useful because it contains the footprint and systematics of the survey, different from just comparing $s^{\rm BAO}_\perp$ that comes from a correlation function based simply on the redshift distribution and a fiducial cosmology.

The final result is the shift parameter $\alpha$, defined as:
\begin{equation}
    \alpha = \frac{D_M(z)}{s_{\perp}^{\rm BAO}}\frac{s_{\perp}^{\rm BAO \, fid}}{D_M^{\rm fid}(z)}
\end{equation}
where $D_M(z)$ is the comoving angular diameter distance. The fiducial model used was from Planck 18 \cite{collaboration2020planck}.

Given our focus on realistic probability distribution functions (PDFs), we calculate the errors by not presuming a Gaussian posterior for the best-fit parameter we use the mode alongside the confidence intervals for plus or minus one standard deviation ($\pm 1\sigma$). Additionally, this approach is applied to determine $D_M(z)$, while incorporating the uncertainty related to photometric redshifts (see Appendix~\ref{ap:error}).

Thus, the final $\alpha$ has asymmetric error bars, as well as its redshift value. In Figure~\ref{fig:bao_post}, we show $\alpha$ for \texttt{ANNz2} (red), \texttt{ANNz2} Gaussian (blue), \texttt{ANNz2} Sp (gray), and \texttt{DNF} (purple). As mentioned in \cite{ferreira2025}, $z_p$ of all samples are compatible and compatible with the DES Y3 (black) result \cite{abbott2022dark}. The main change is the shift parameter, we see that \texttt{ANNz2} Gaussian is compatible with DES Y3 and the mock estimation (salmon). The main conclusion is that the Gaussian sample selection is closer to the fiducial result and seems to follow the DES Y3 result. In other words, this sample was shown to be the best representation of the spectroscopic reference set \cite{ferreira2025} and shows a stronger correlation to the \textit{Planck 18} $\Lambda$CDM scenario, while other samples seem to deviate from CMB results especially \texttt{DNF} with the lower $\alpha$. However, once all available PDFs are taken into account, the apparent agreement between \texttt{ANNz2} and \texttt{DNF} suggests that spurious PDFs can substantially bias and shift the inferred cosmological constraints.

\section{Transverse correlation function from PDF selection}\label{sec:cosmology_model}

 The challenge of photometric surveys is to account for its photometric systematics. \cite{chan2022clustering} proposed a method that uses the angular correlation function, $w(\theta, z_p,z_p')$, in the 3D power spectrum model. The idea is simple: there is a relation between the function and the power spectrum given a redshift distribution of a redshift bin. 

The expression for a general cross-correlation angular function between two bins with photo-z $z_p$ and $z_p'$ is
\begin{equation}
    w(\theta, z_p,z_p') = \int \mathrm{d} z \phi(z|z_p)\int \mathrm{d} z' \phi(z'|z_p')\int \frac{\mathrm{d} \mathbf{k}}{(2\pi)^3} P(\mathbf{k},z,z') e^{i\mathbf{k} \cdot [\mathbf{r}(z)-\mathbf{r}' (z')]}.
\end{equation}
$\phi(z|z_p)$ is the redshift distribution of a redshift bin, it is a function of the bin's PDF. This is the distribution function of a group of galaxies with a true redshift $z$, from a spectroscopic sample, given that we have information of $z_p$. $\phi(z|z_p)$ is written as:
\begin{equation}\label{eq:phi}
    \phi(z|z_p) = f(z|z_p) \frac{\bar{n}(z)}{\bar{n}_p(z_p)},
\end{equation}
where $f(z|z_p)$ is the conditional distribution of redshift based on the $PDF$ values that correspond to an expected spectroscopic redshift distribution $n_{DES}(z_s)$. When we compute $f(z|z_p)$\footnote{The code for this computation is found at \url{https://github.com/psilvaf/bao_pz}.}, we are considering all the matched galaxies we used to construct \texttt{Pz Cats} (see companion paper \cite{ferreira2025}) instead of only selecting the VIMOS Public Extragalactic Redshift Survey (\texttt{VIPERS}) sample as done in \cite{chan2022dark}. $\bar{n}(z)$ is the distribution of galaxies with expected $z$, and $\bar{n}_p(z_p)$ is the distribution of galaxies with a corresponding $z_p$.

Taking into account the multipole expansion of $P(\mathbf{k},z,z')= \sum_{even\text{ }\ell} P_\ell(k,z,z')\mathcal{L}_\ell(\mathbf{\hat{k}}\cdot \mathbf{\hat{r}}_{||})$, where $\mathcal{L}_\ell$ is the Lengendre polynomial. The angular cross-correlation between bins $(z_{p})$ and $(z_{p}')$ can be written as 
\begin{equation}\label{eq:crosscorr}
    w(z_p,z_p',\theta)= \sum_\ell i^\ell \int \mathrm{d} z \phi(z|z_p) \int \mathrm{d} z' \phi(z'|z_p') \mathcal{L}_\ell (\hat{\mathbf{r}} \cdot \mathbf{\hat{r}}_{||}) \int \frac{\mathrm{d}kk^2}{2\pi^2}j_\ell(kr) P_\ell(k,z,z'),
\end{equation}
where $\theta$ represents the angular separation between bins $z_p$ and $z_p'$, $\mathbf{\hat{r}}_\parallel$ is the unitary vector parallel to the line-of-sight, and $\mathbf{\hat{r}}$ is the unitary vector parallel to the comoving distance from the observer to the source.

Instead of writing a function $\xi(s_\perp)$, as proposed by \cite{chan2022clustering}, by retrieving the least information in the line of sight from the cross-correlation between the chosen bins ($w(\theta(\mu),z_p,z_p')$), we make use of the function $w(\theta,z_p,z_p')$ choosing the angle where the bins are perpendicular to the observer's line of sight, specifically at $\theta=180^o$, as the normalisation factor. When $\theta=180^\circ$, it represents the uncorrelated scenario. This normalisation retains its interpretation within the definition of the correlation function, where we compare the correlation that includes the BAO feature to the scenario where no such feature is present. We retain all possible angles rather than excluding those near the line-of-sight as suggested by \cite{chan2018bao}, thereby maintaining the transverse relationship that is central to our study.

The monopole $\ell=0$ leads to:
\begin{equation}
    w(\theta,z_p,z_p')=\int \dd z f(z|z_p) \int \dd z' f(z'|z_p') \int \frac{\dd k \, k^2}{2 \pi^2}P_0(k,z,z'),
\end{equation}
then we can interpret the results for $\xi(s_\perp)$ and $\xi(s_\parallel)$.

    \begin{equation}
    \frac{w(\theta,z_p,z_p')}{w(\theta \simeq 0^o,z_p,z_p')} \, \longrightarrow \, \xi_\parallel(z_p,z_p')=\xi_\parallel(\Delta z_p)
\end{equation}
\begin{equation}
    \frac{w(\theta,z_p,z_p')}{w(\theta \simeq 180^o,z_p,z_p')} \, \longrightarrow \, \xi_\perp(z_p,z_p')=\xi_\perp(\Delta z_p)
\end{equation}

The interpretation of this function is simply counting pairs of objects separated by $\Delta z_p$, the highest correlation is restricted to the closest bin. Then the correlation decreases as one increases $\Delta z$. The BAO feature should appear as a bump at a preferable separation between the bins according to the comoving distance between the two bins:
\begin{equation}
    d_c = \frac{c}{H_0}\int_{z_p}^{z_p'}\frac{\dd z}{E(z)}.
\end{equation}
 From \cite{takahashi2012revising}, we write the relation between the angular and three-dimensional power spectrum. For that, there is a transformation between the wave-number $k$ into the multipole $\ell$, this requires the use of a fiducial cosmology to convert redshift into comoving distance $\tilde{\chi}$, which is the best fit Planck 18.
\begin{equation}
C_\ell = \int \dd z f(z|z_p) P\left(k=\frac{\ell}{\tilde{\chi}(z)};z\right)
\end{equation}
We use $f(z|z_p)$ instead of $\phi(z|z_p)$, because we want retrieve information purely from the galaxies' PDFs, while $\phi(z|z_p)$ imposes a smoothing effect from $\bar{n}(z)$.

Next, we find the angular correlation function through a Fourier transform of $C_\ell$:
\begin{equation}
    w(\theta) = \sum_\ell \frac{(2\ell+1)}{4 \pi } C_\ell P_\ell(\cos{\theta}).
\end{equation}
Finally, to get $\xi(s_\perp)$ we can simply integrate the equation above using the bins photo-z distribution 
\begin{equation}\label{eq:final_model}
    \xi(s_\perp) = \int \dd z_p \int \dd z_p' \frac{w(\theta,z_p,z_p')}{w(\theta \simeq 180^o,z_p,z_p')}.
\end{equation}

\begin{figure}
    \centering
    \begin{subfigure}{0.45\linewidth}
        \centering
        \includegraphics[width=\linewidth]{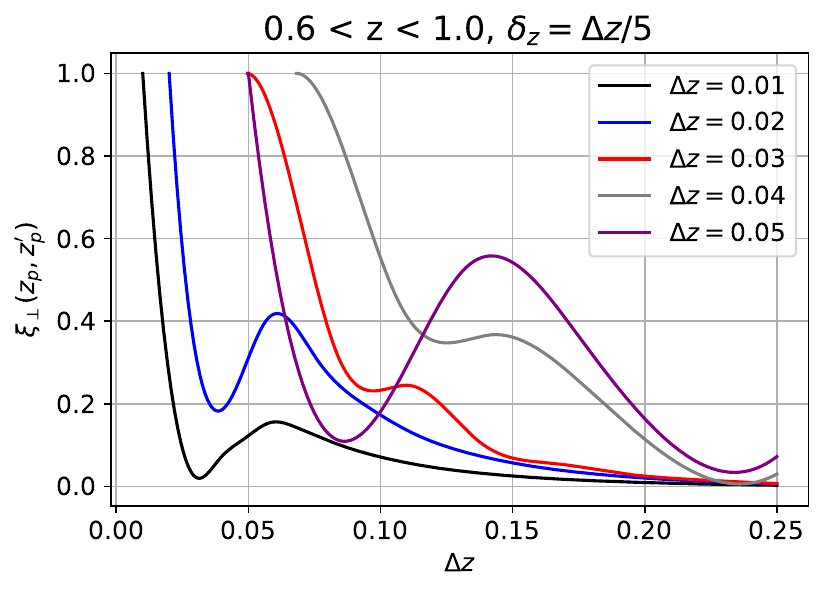}
        \caption{Gaussian Bins with standard deviation $\delta_z = \Delta z/5$.}
        \label{fig:wide_bins}
    \end{subfigure}
    \hfill
    \begin{subfigure}{0.45\linewidth}
        \centering
        \includegraphics[width=\linewidth]{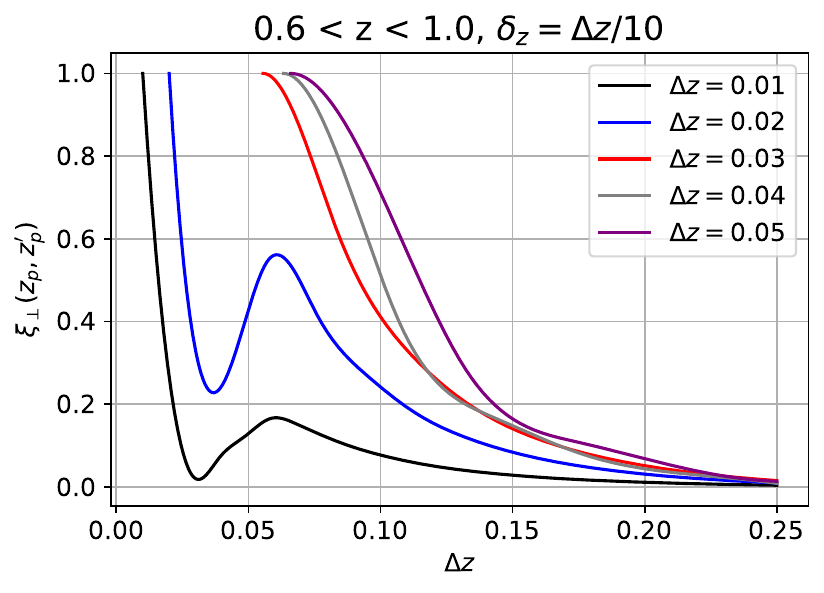}
        \caption{Gaussian Bins with standard deviation $\delta_z = \Delta z/10$.}
        \label{fig:thin_bins}
    \end{subfigure}
    \caption{Comparison of wide and thin bins. Separating the bins by fixed $\Delta z$: \textit{black}: 0.01, \textit{blue}: 0.02, \textit{red}: 0.03, \textit{gray}: 0.04, \textit{purple}: 0.05.}
    \label{fig:bins_comparison}
\end{figure}

We tested how the separations between the bins $\Delta z$ and the bins width $\delta_z$ change the BAO. In Figure~\ref{fig:bins_comparison}, we show the relation of crossing wide or thin bins. The test for a survey in which the matched redshift distribution is from the samples we chose, and each bin distribution ($f(z|z_p)$ is described by a Gaussian whose standard deviation is $\delta_z=\Delta z /5$. We know that the BAO feature has different scales depending on the redshift of the sample. Here, we cross bins, so the BAO feature is placed at a particular $\Delta z$ when there is an optimal number of adjacent bins so that the correlation between them is related to the BAO signal.

In the first panel, figure~\ref{fig:wide_bins}, the lower separation gives the feature at later times than for bigger $\Delta z$. This is clear for the sequential colours black, blue, red, and grey. In figure~\ref{fig:thin_bins}, we force thin bins $\delta_z=\Delta z/10$, the dependence on correlation is evident, the most distant bins are very uncorrelated, so there is no BAO. Lastly, we notice that the purple line in figure~\ref{fig:wide_bins} is shifted compared to the grey one, but we expect the BAO scale to change monotonically. The reason for that is simply a relation of overlapping bins, for wider bins, an extra correlation between neighbours increases the signal and forces a shift toward lower separation.
\subsection{The inclusion or not of line-of-sight terms}\label{ap:mu}

\begin{figure}
    \centering
    \includegraphics[width=\linewidth]{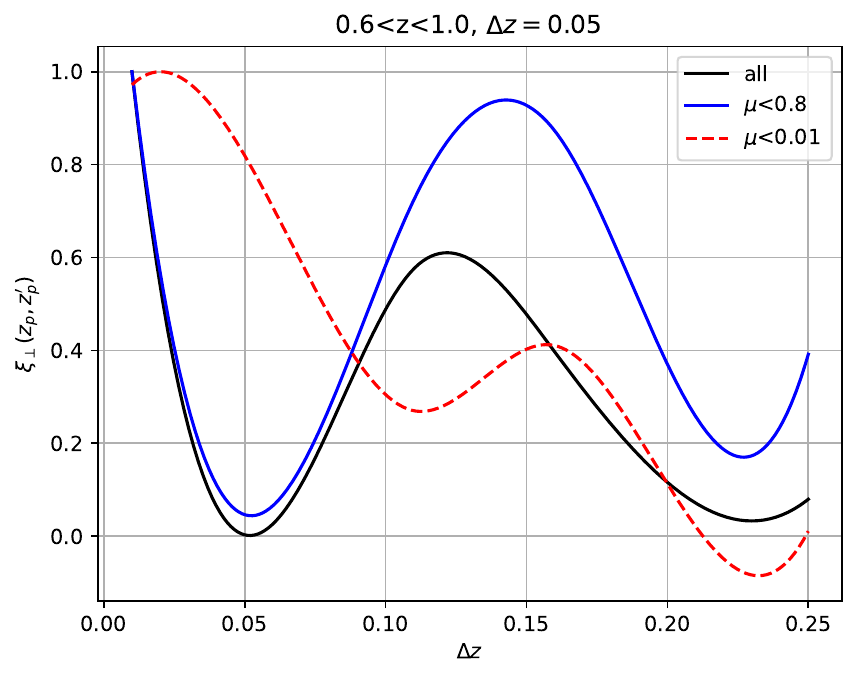}
    \caption{Comparison of different $\xi(s_\perp)$ by including all the possible cross-correlation (black), when $\mu<0.8$ (blue), and when $\mu<0.01$ (dashed red line).}
    \label{fig:mu}
\end{figure}

We test the exclusion of the angular correlation function values too close to the line-of-sight instead the full acceptance we propose. The idea is to select $\mu$ up to a $\tilde{\mu}$ as in
\begin{equation}
        \xi_\perp(\Delta z)=\int dz_p \int dz_p'\frac{w(\theta(\mu<\tilde{\mu}),z_p,z_p')}{w(\theta \simeq 180^o,z_p,z_p')}.
\end{equation}
The result can be found in Figure~\ref{fig:mu}, in blue is our approach, in black and red we chose $\mu<0.8$ and $\mu<0.01$, respectively. When $\mu$ can be of any value, the BAO position is widened and the correlation function amplitude increases. For $\mu<0.8$, we still have the projection effect issue with smaller amplitude and width. Finally, the least line-of-sight scenario looses signal, but shifts the BAO to higher $\Delta z$.

\section{Results with bin crossing}\label{sec:results}
\subsection{The four estimators model results}

\begin{figure}[htbp]
    \centering
    \begin{subfigure}[b]{0.45\textwidth}
        \centering
        \includegraphics[width=\linewidth]{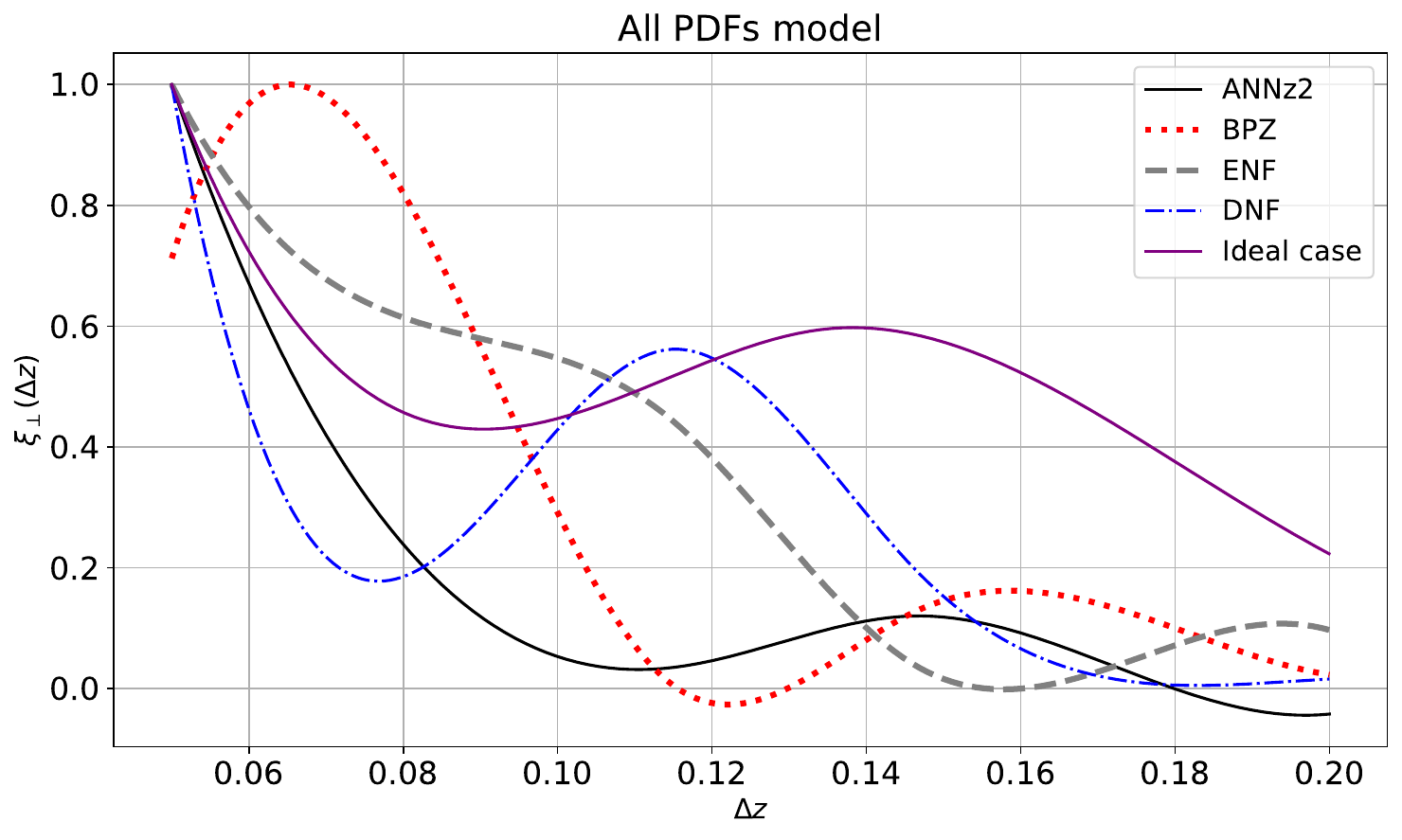}
        \caption{Models with criteria 1}
        \label{fig:dif_models1}
    \end{subfigure}
    \hfill
    \begin{subfigure}[b]{0.45\textwidth}
        \centering
        \includegraphics[width=\linewidth]{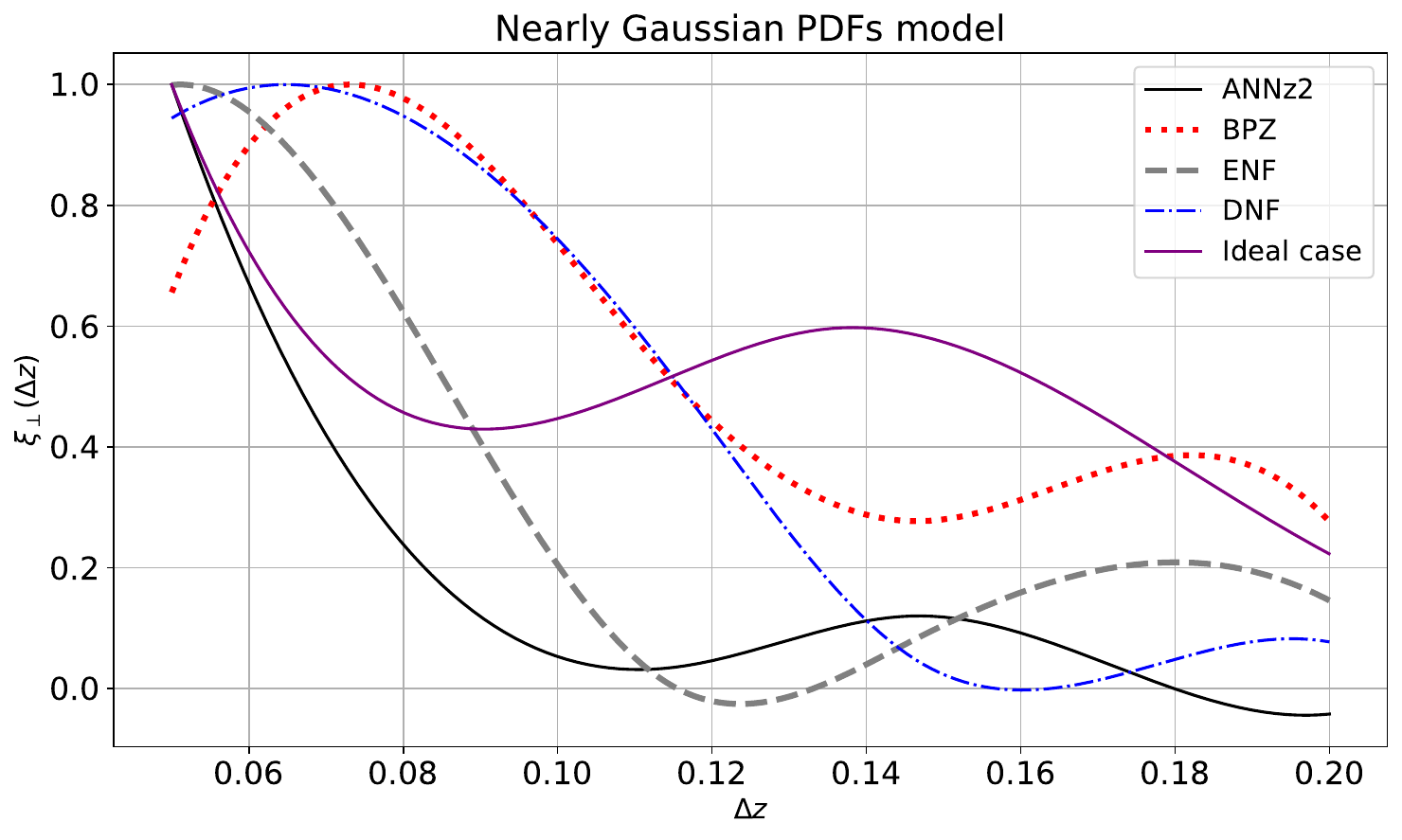}
        \caption{Models with criteria 2}
        \label{fig:dif_models2}
    \end{subfigure}
    
    \begin{subfigure}[b]{0.5\textwidth}
        \centering
        \includegraphics[width=\linewidth]{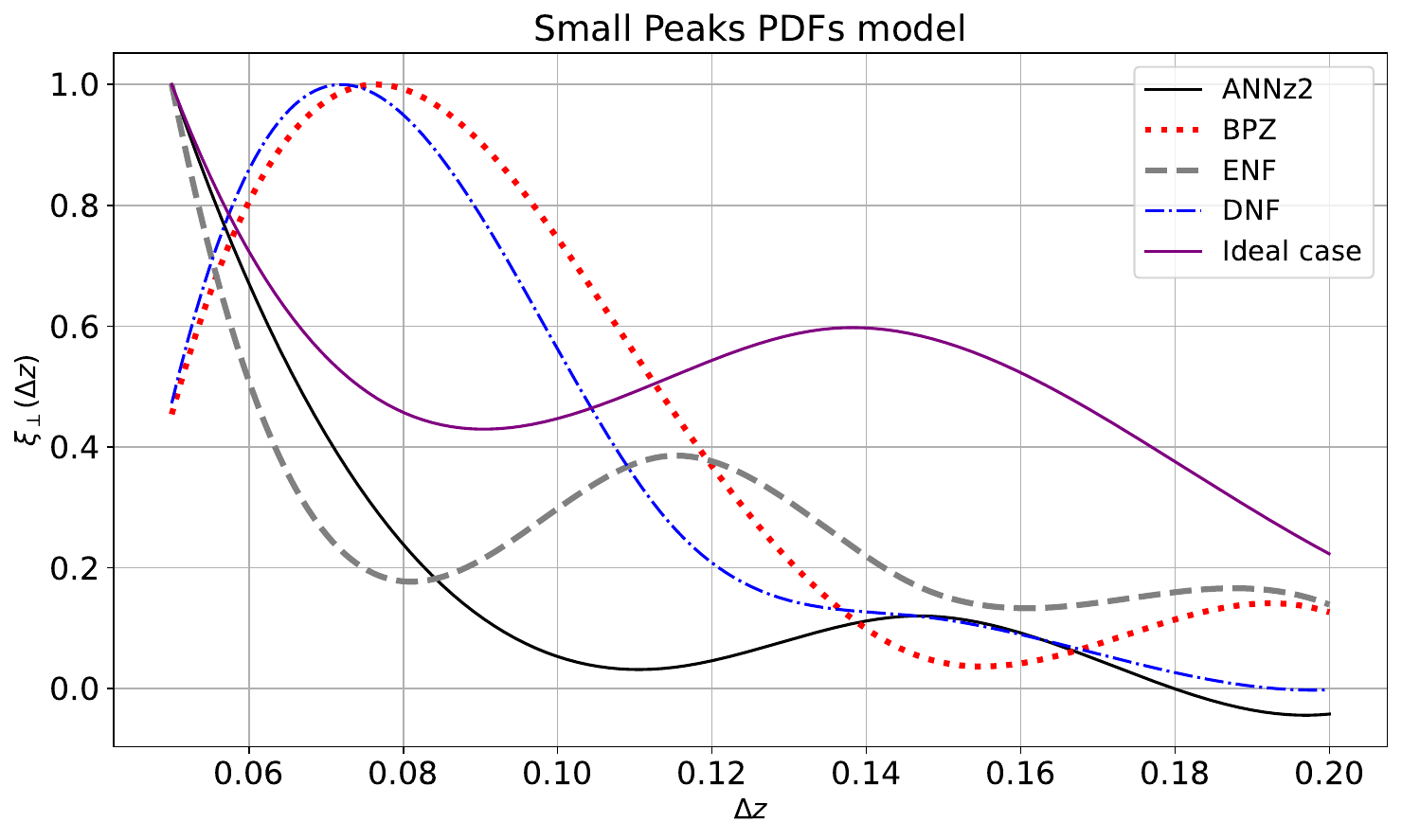}
        \caption{Models with criteria 3}
        \label{fig:dif_models3}
    \end{subfigure}
    
    \caption{Comparison of different $\xi(s_\perp)$ from different models (1- full sample, 2 - Gaussian, 3 - Small Peaks) with PDFs selection criteria. \textit{black}: \texttt{ANNz2} model, \textit{red}: \texttt{BPZ}, \textit{gray}:\texttt{ENF}, \textit{blue}: \texttt{DNF} and \textit{purple}: ideal case (Perfectly Gaussian $\phi(z|z_p)$).}
    \label{fig:all_models}
\end{figure}

Given Eq.~(\ref{eq:final_model}), we show the results using each estimator and its corresponding sample cut in Figure~\ref{fig:all_models}. For all cases, we set the Planck 18 \cite{abbott2018dark} $\Lambda$CDM parameters to compute $P_0(k,z,z')$, for that we use \texttt{CAMB} \cite{lewis2011camb}. The first plot Fig.~\ref{fig:dif_models1} we show the case with all the galaxies included in the survey to select the function $\phi(z|z_p)$. We see that \texttt{ANNz2} has the most accurate BAO position compared to the Ideal case where $\phi(z|z_p)$ is a Gaussian distribution (in purple also with Planck 18 $\Lambda$CDM parameters). 

Figure~\ref{fig:dif_models1} shows the case where we make use of all the galaxies available. Compared to the Ideal case, \texttt{ANNz2} (in black) has the BAO feature closer to the purple line. \texttt{DNF} has a higher amplitude matching the ideal case, but the BAO feature is shifted toward smaller scales. \texttt{BPZ} seems to place the BAO signal on higher scales, while \texttt{ENF} does not have the signal at all. When we choose to select only the PDFs close to a Gaussian distribution, Figure~\ref{fig:dif_models2} shows that \texttt{BPZ} and \texttt{ENF} have the feature with 2$\sigma$ with the Ideal case, \texttt{ENF} has it located at higher scales, \texttt{DNF} does not present the feature, and \texttt{ANNz2} matches within 1$\sigma$ with the expected result. Lastly, in Fig.\ref{fig:dif_models3}, when the least noisy PDFs are selected, \texttt{ENF} and \texttt{ANNz2} have the BAO signal, the first on a smaller scale with a bigger signal and the second on a larger scale. \texttt{DNF} and \texttt{BPZ} do not have the feature.

This preliminary result indicates two things. First, the main photo-z estimator of the DES Collaboration is efficient when there are enough statistics, however, once there is a reduction in the number of objects, the feature disappears. Second, the PDF influence in the BAO position is shifted depending on the estimator, except for \texttt{ANNz2} which is consistent within the three cases. From Figure \ref{fig:bins_comparison}, we know that shifts are strongly related to the $z_{\rm eff}$, because the algorithms we used resulted in compatible $z_{\rm eff}$ the different BAO features are the result in the change of statistics and photo-z precision.

\subsection{Bayes Factor}

To support our previous arguments, we must test a set of models with different cosmological parameters and compare the likelihoods between different $\xi_\perp(\Delta z_p)$ models. We choose a particular model as the fiducial scenario, $\phi(z_i|z_{ip})$ is Gaussian and with the same amplitude for all $i$-th bins, this is never possible in reality. 

Considering four distinct photo-z estimators and three potential sample selections, we examined two cosmological parameters by varying them within a normal distribution, where the standard deviation was twice that from the Planck 18 results: the Hubble constant $H_0$ [$\mathcal{N}(67.27,1.2)$], the baryonic density parameter $\Omega_b\, h^2$ [$\mathcal{N}(0.02236,0.0003)$], the wCDM equation of state parameter $w_0$[$\mathcal{N}(-0.990,0.001)$], and the baryonic density again $\Omega_b\, h^2$, along with the galaxy bias parameter $\nu$[$\mathcal{N}(0.50,.01)$] applied to a bias redshift relation 
\begin{equation}
    b(z)=(1+z)^\nu.
\end{equation}

The chi-squared distribution for each correlation function $\xi^A(\Delta z_\iota,\vec{\theta})$, for a given estimator $A$ varying the parameters $\vec{\theta}$ and compared to the expected result of fiducial ideal case $fid$, which we chose the absolute value of the 50th percentile of that distribution $|\xi^{fid_{p50}}(\Delta z_\iota,\vec{\theta})|$, is written as:

\begin{equation}
    \chi^2_A(\vec{\theta}) = \sum_{\iota} \frac{(\xi^{A}(\Delta z_\iota, \vec{\theta}) - \xi^{fid}(\Delta z_\iota, \vec{\theta}))^2}{|\xi^{fid_{p50}}(\Delta z_\iota, \vec{\theta})|} + \sum_{\iota \neq \kappa} \frac{(\xi^{A}(\Delta z_\iota, \Delta z_\kappa, \vec{\theta}) - \xi^{fid}(\Delta z_\iota, \Delta z_\kappa, \vec{\theta}))^2}{|\xi^{fid_{p50}}(\Delta z_\iota, \Delta z_\kappa, \vec{\theta})|},
\end{equation}

where the first term represents the original contribution from individual redshift separations indexed by $\iota$, and the second term accounts for correlations between different bin separations $\iota$ and $\kappa$. The cross-bin terms arise from potential statistical dependencies or systematic effects that couple different redshift intervals.

In practice, however, the contribution from cross-bin correlations is often negligible compared to the dominant diagonal terms. This occurs for two main reasons. First, if the correlation between $\Delta z_\iota$ is weak because the correlation between bins already weakens the signal. The difference $\xi^{A}(\Delta z_\iota, \Delta z_\kappa, \vec{\theta}) - \xi^{fid}(\Delta z_\iota, \Delta z_\kappa, \vec{\theta})$ becomes very small for $\iota \neq \kappa$, suppressing the cross-term. Second, the effective variance $|\xi^{fid_{p50}}(\Delta z_\iota, \Delta z_\kappa, \vec{\theta})|$ for off-diagonal elements tends to be larger than the diagonal terms, further reducing their relative weight in the chi-squared sum.

As a result, the expression can typically be well-approximated by the original simplified form that only considers individual bin contributions:

\begin{equation}
    \chi^2_A(\vec{\theta}) \approx \sum_{\iota} \frac{(\xi^{A}(\Delta z_\iota, \vec{\theta}) - \xi^{fid}(\Delta z_\iota, \vec{\theta}))^2}{|\xi^{fid_{p50}}(\Delta z_\iota, \vec{\theta})|}.
\end{equation}

This approximation remains valid as long as the cross-bin correlations do not contain significant cosmological information or introduce substantial systematic effects that need to be explicitly modeled in the analysis.

Next, the likelihood function relation with the Chi-squared function is:
\begin{equation}
    \mathcal{L}_A(\vec{\theta}|A) \propto \sum_{\vec{\theta}}\exp{-\sum_\iota [\chi^2_\iota(\vec{\theta})]_A}.
\end{equation}
The likelihood $\mathcal{L}_A$ gives the probability distribution of finding the parameters $\vec{\theta}$ given the model $A$.  

\begin{table}
    \centering
    \begin{tabular}{c|c}
        \hline
        $B_{AA'}$ & Evidence \\
        \hline
         $1\leq B_{AA'}<3$&  Weak \\
         $3\leq B_{AA'}< 20$ & Definite \\
         $20 \leq B_{AA'} < 150$ & Strong \\
         $150 \leq B_{AA'}$ & Very Strong \\
         \hline
    \end{tabular}
    \caption{Jeffreys' scale}
    \label{tab:jeff}
\end{table}

Finally, the Bayes Factor $B_{AA'}$ comparing two different models is the ratio of their respective marginal likelihoods integrated for all parameters. 
\begin{equation}
    B_{AA'}=\frac{\mathcal{L}_A}{\mathcal{L}_{A'}}.
\end{equation}
This equation tells the evidence of the model $A$ against the model $A'$. Whether the evidence is sufficient or not, we will follow the interpretation of the Jeffreys' scale based on reference \cite{nesseris2013jeffreys} which we show in table~\ref{tab:jeff}.

\begin{figure}
    \centering
    \includegraphics[width=\linewidth]{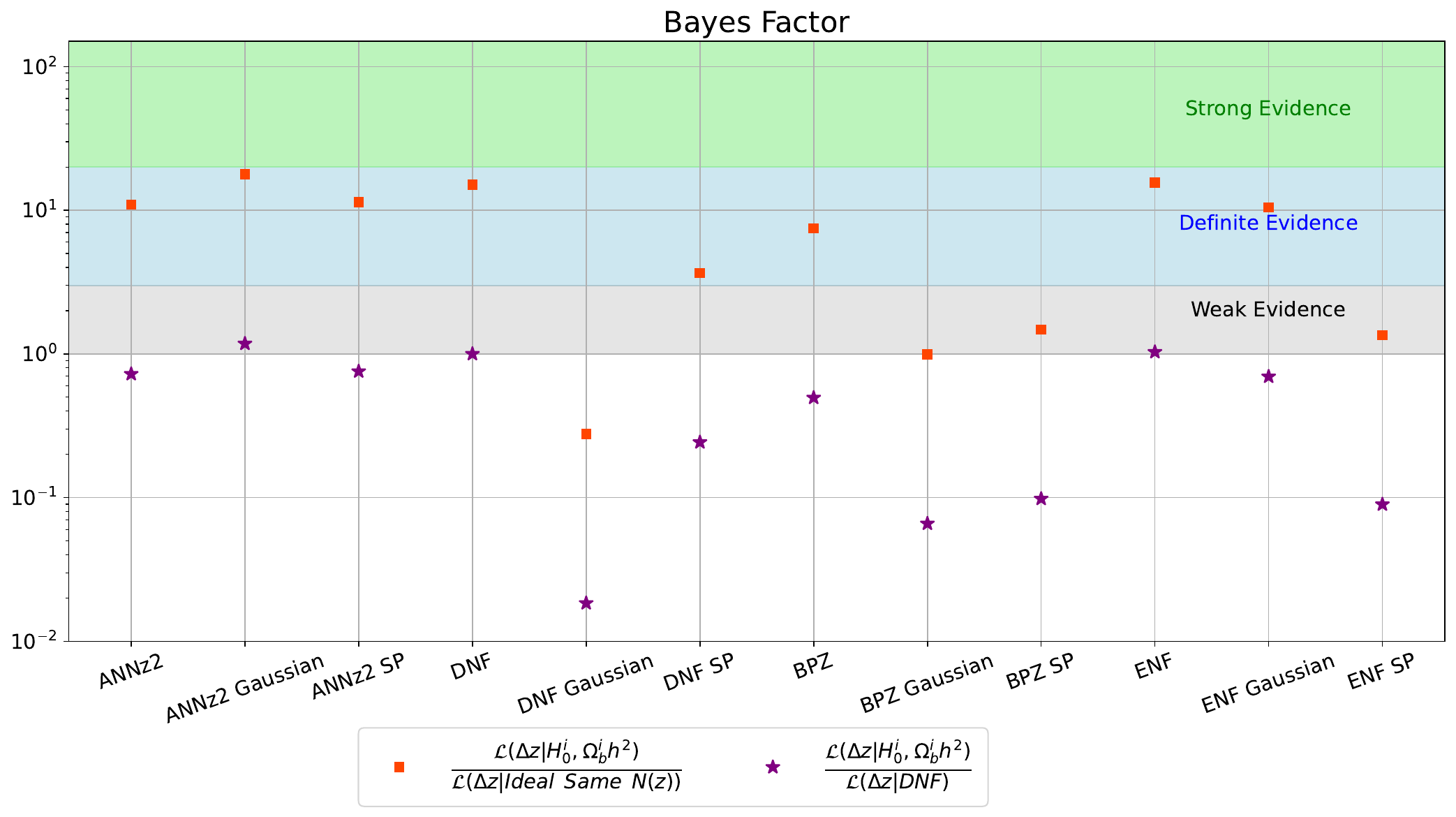}
    \caption{Bayes factor when varying $H_0$ and $\Omega_m h^2$ from the observed PDFs of each photo-z estimator. The \textit{gray} region represents weak evidence, the \textit{blue} region represents definite evidence, and the \textit{green} region the strong one. \textit{Red} squares represents comparison with the ideal resulting correlation function and \textit{purple} stars comparing with the \texttt{DNF} distributions.}
    \label{fig:bayes}
\end{figure}

In Figure~\ref{fig:bayes}, we show the Bayes factor of each estimator compared to another model. The gray colour represents the region where the evidence is weak, blue is for definite evidence, and green is for strong evidence. We tested two different models to compare, one is an ideal scenario where $f(z|z_p)$ is Gaussian (red square) and the other compares the likelihoods with the DES Collaboration main photo-z estimator \texttt{DNF} (purple star). 
We see that the estimators \texttt{DNF} Gaussian and SP, \texttt{ENF} Gaussian and SP, all \texttt{BPZ} samples, and \texttt{ANNz2} SP have no evidence against a model considering \texttt{DNF} full survey sample. The exceptions are only \texttt{ANNz2} Gaussian and \texttt{ENF}. 
Compared to the expected model (red squares), \texttt{ANNz2} Gaussian has the best performance. The others remain in the same pattern as the last paragraph. This results match the performance of the photo-z estimator we found in the companion paper \cite{ferreira2025}. \texttt{ANNz2} is the best estimator for the available spectroscopic sample we had, which means the PDF's shape is relevant for BAO analysis and the PDF for different estimators is also appropriate. \texttt{DNF} is capable of improving performance when there are enough statistics, whenever we seek sample cuts based on the PDFs this estimator loses performance.
\texttt{BPZ} also improves with enough statistics, even though there are many galaxies with Gaussian PDFs, these distributions may be showing the wrong redshift result, we showed that this is true in the companion paper \cite{ferreira2025}.

\begin{figure}
    \centering
    \includegraphics[width=\linewidth]{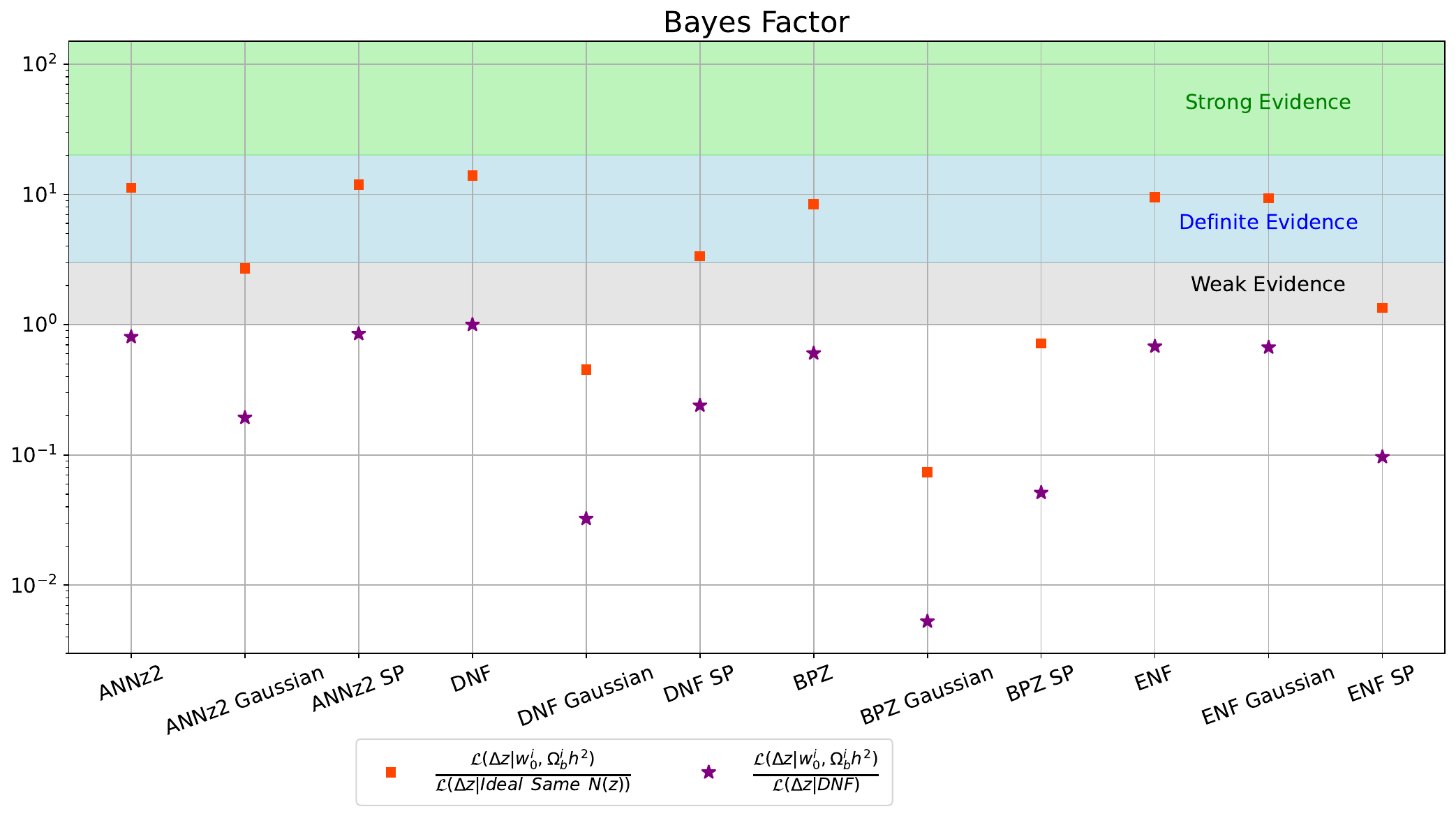}
    \caption{Bayes factor when varying $w_0$ and $\Omega_b h^2$ from the observed PDFs of each photo-z estimator. The \textit{gray} region represents weak evidence, the \textit{blue} region represents definite evidence, and the \textit{green} region the strong one. \textit{Red} squares represents comparison with the ideal resulting correlation function and \textit{purple} stars comparing with the \texttt{DNF} distributions.}
    \label{fig:bayes2}
\end{figure}

Now, the second set of parameters was used, without considering $H_0$, a parameter that is not directly measured by BAO. In Figure~\ref{fig:bayes2}, we vary $\Omega_b$ and $w_0$, the change was significant to \texttt{ANNz2} Gaussian, which changed from definite evidence to weak evidence compared to the expected model (red squares). For this case, \texttt{DNF} performs better than \texttt{ANNz2}, \texttt{ANNz2} SP remains with definite evidence, and \texttt{BPZ} Gaussian loses more information. 

\begin{figure}
    \centering
    \includegraphics[width=\linewidth]{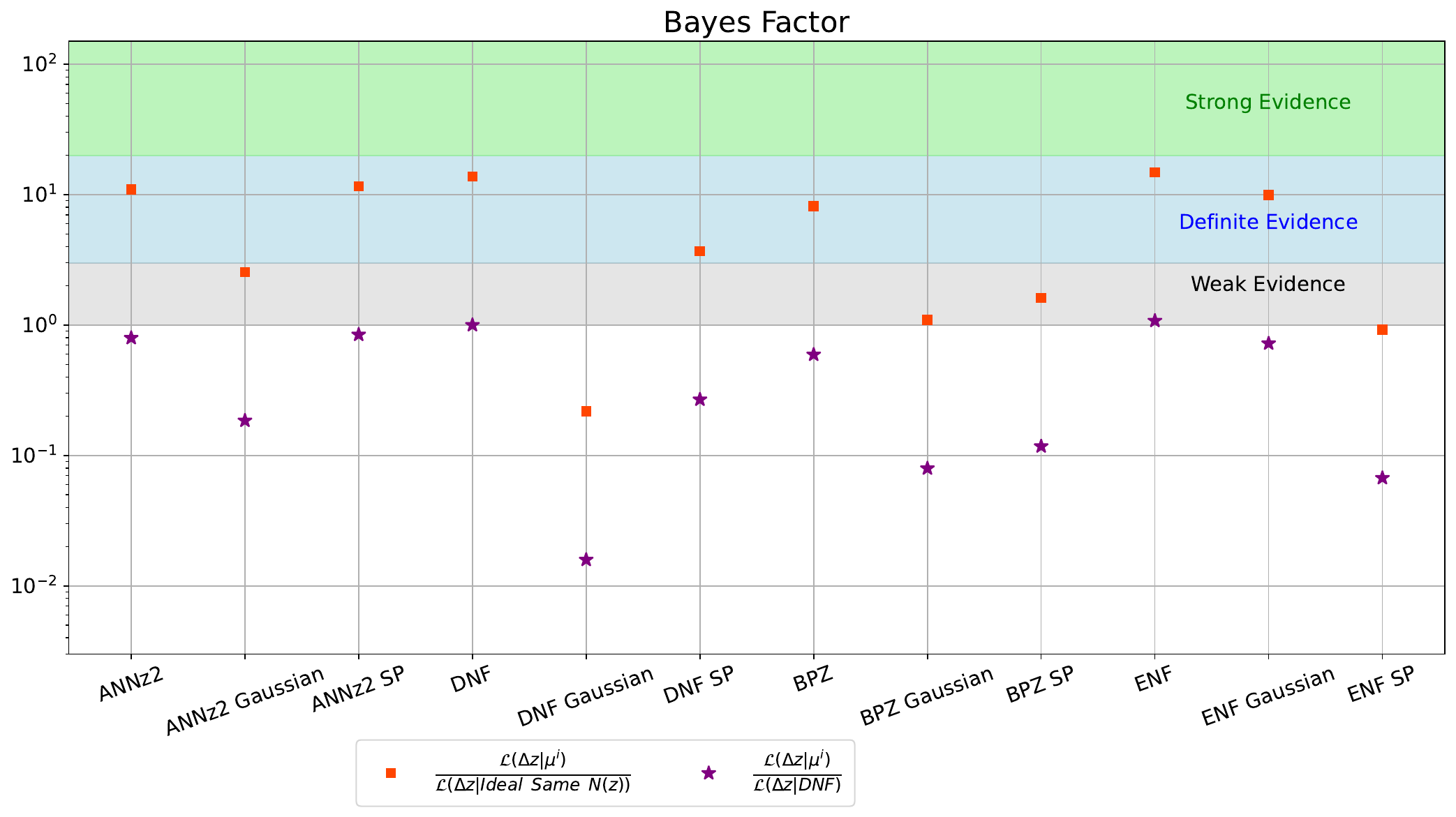}
    \caption{Bayes factor when varying the bias parameter $\mu$ for a bias relation $b(z)=(1+z)^\mu$ from the observed PDFs of each photo-z estimator. The \textit{gray} region represents weak evidence, the \textit{blue} region represents definite evidence, and the \textit{green} region the strong one. \textit{Red} squares represents comparison with the ideal resulting correlation function and \textit{purple} stars comparing with the \texttt{DNF} distributions.}
    \label{fig:bayes3}
\end{figure}

Finally, we compared the algorithms perfomances with respect to galaxy bias. Figure~\ref{fig:bayes3} shows the change in results. The main difference in performance apears to \texttt{ENF} SP, compared to the expected model it has indefinite evidence. \texttt{BPZ} Gaussian has now weak evidence against the fiducial model, once again, in agreement to Fig.~\ref{fig:bayes} in which the parameter $\Omega_b$. The others maintain the patterns described in the last paragraph.

In the end, we found that for the same fiducial model, with small fluctuations, the photo-z influences the constraints. When the PDFs are relatively smooth and symmetric, the evidence tends to be stronger, except in cases with significant statistical loss. We advise future photo-z survey collaborations to consider including the cosmogical inference with more than one photo-z estimation algorithm; this would allow a deeper understanding of systematic influence for different parameters.

\texttt{DNF} is a good estimator when the number of objects is significant enough to compensate for the imperfections in each galaxy's PDF. In the companion paper \cite{ferreira2025}, we noticed that \texttt{DNF}'s performance relies strongly on a highly populated training set. We conclude that using the training set we could access for this estimator is not ideal, the preferable choice is \texttt{ANNz2}. 

\section{Summary}\label{sec:conclusion}
In this paper, we tested the influence of the redshift PDFs in finding the BAO feature. First, we selected the best PDFs for four different photo-z estimators: \texttt{ANNz2}, \texttt{BPZ}, \texttt{ENF}, and \texttt{DNF}. We used the full sample to select the nearly Gaussian PDFs and the least noisy ones. As expected some resulting samples lose statistics significantly, \texttt{ENF} Gaussian has the smallest number of galaxies. 

The first test was done with a model fitting using MCMC to constrain the parameters of a polynomial based on \cite{sanchez2013precise}. From the estimated correlation function from the data samples, we found the best-fit to the polynomial function. Only \texttt{ANNz2} and its sub-samples and \texttt{DNF} full samples were capable of finding the BAO signal directly by estimating the projected correlation function $\xi(s_\perp)$. We computed the shift parameter $\alpha$ for these samples and found that \texttt{ANNz2} Gaussian sample selection is closer to the fiducial result and seems to follow the DES Y3 result as well. Overall, considering all available PDFs, the close agreement between \texttt{ANNz2} and \texttt{DNF} indicates that spurious PDFs can significantly bias and shift the resulting cosmological constraints.

Motivated by this first result, we adapted the model from \cite{chan2022clustering} to get the transverse correlation function $\xi_\perp(z_p)$ by getting the bin pairs transversal to each other using \texttt{CAMB}. The kernel window function is the $f(z|z_p)$ which is the selection of the PDF value when the photometric redshift is nearly the same as the spectroscopic redshift estimated by the matched spectroscopic sample.  

As expected by the theoretical background, the width of the bins influences the position of the BAO feature, this consistency check confirms our proposed method is reliable. For higher redshift bins, the width must be enough so that the neighbouring shells are correlated. The thinner the bin is, the higher redshift shells are not correlated and will not show the BAO signal.

With the realistic scenario, we tested $\xi_\perp$ for each estimator with the $\Lambda$CDM model. \texttt{DNF} is efficient when there are enough statistics, once there is a reduction in the number of objects, the feature disappears. The PDF influence in the BAO position is shifted depending on the estimator, except for \texttt{ANNz2} which is consistent within the three cases. 

After that, we compared the Bayes Factor with a set of models varying $H_0$, $\Omega_b \, h^2$, $w_0$, and galaxy bias. We used two models as the reference models, one is the $\Lambda$CDM and $f(z|z_p)$ are Gaussian, and the other is the \texttt{DNF} result. Again, statistical loss plays a significant role in the evidence, the only estimator to succeed in all the PDF selections was \texttt{ANNz2}, where the Gaussian distributions showed more substantial evidence either compared to \texttt{DNF} or compared to the ideal scenario. So we concluded that the shape of the galaxy redshift PDF could shift the BAO feature position either by including the PDF in the model or not for compatible $z_{\rm eff}$. We also learnt that given the same spectroscopic sample, \texttt{ANNz2} is the best photo-z estimator for BAO analysis considering the realistic probability distribution function of each galaxy. Ultimately, it was determined that for an identical fiducial model experiencing minor variations, the photometric redshift impacts the constraints. Stronger evidence is observed when the probability distribution functions are predominantly smooth and symmetric, except where considerable statistical decay occurs. We recommend that future photometric redshift survey collaborations contemplate incorporating cosmological inference using multiple photo-z estimation algorithms, as this could enhance comprehension of systematic effects on various parameters.

\section{Code availability}
The particular software packages used in this work will be accessible at \url{https://github.com/psilvaf/bao_pz}, \url{https://github.com/psilvaf/cat_org}, and \url{https://github.com/psilvaf/mock_gen}. 

\section*{Acknowledgements}
This work made use of the CHE cluster, managed and funded by COSMO/CBPF/MCTI, with financial support from FINEP and FAPERJ grant E-26/210.130/2023, and operating at the Javier Magnin Computing Center/CBPF.

PSF thanks Brazilian funding agency CNPq for PhD scholarship GD 140580/2021-2. RRRR thanks CNPq for partial financial support (grant no. $309868/2021-1$).

The authors would like to thank the anonymous referee who provided useful and detailed comments.

The authors thank Juan De Vicente (Centro Investigaciones Energéticas, Medioambientales y Tecnológicas) for providing the \texttt{DNF} estimator code.

This project used public archival data from the Dark Energy Survey (DES). Funding for the DES Projects has been provided by the U.S. Department of Energy, the U.S. National Science Foundation, the Ministry of Science and Education of Spain, the Science and Technology FacilitiesCouncil of the United Kingdom, the Higher Education Funding Council for England, the National Center for Supercomputing Applications at the University of Illinois at Urbana-Champaign, the Kavli Institute of Cosmological Physics at the University of Chicago, the Center for Cosmology and Astro-Particle Physics at the Ohio State University, the Mitchell Institute for Fundamental Physics and Astronomy at Texas A\&M University, Financiadora de Estudos e Projetos, Funda{\c c}{\~a}o Carlos Chagas Filho de Amparo {\`a} Pesquisa do Estado do Rio de Janeiro, Conselho Nacional de Desenvolvimento Cient{\'i}fico e Tecnol{\'o}gico and the Minist{\'e}rio da Ci{\^e}ncia, Tecnologia e Inova{\c c}{\~a}o, the Deutsche Forschungsgemeinschaft, and the Collaborating Institutions in the Dark Energy Survey.
The Collaborating Institutions are Argonne National Laboratory, the University of California at Santa Cruz, the University of Cambridge, Centro de Investigaciones Energ{\'e}ticas, Medioambientales y Tecnol{\'o}gicas-Madrid, the University of Chicago, University College London, the DES-Brazil Consortium, the University of Edinburgh, the Eidgen{\"o}ssische Technische Hochschule (ETH) Z{\"u}rich,  Fermi National Accelerator Laboratory, the University of Illinois at Urbana-Champaign, the Institut de Ci{\`e}ncies de l'Espai (IEEC/CSIC), the Institut de F{\'i}sica d'Altes Energies, Lawrence Berkeley National Laboratory, the Ludwig-Maximilians Universit{\"a}t M{\"u}nchen and the associated Excellence Cluster Universe, the University of Michigan, the National Optical Astronomy Observatory, the University of Nottingham, The Ohio State University, the OzDES Membership Consortium, the University of Pennsylvania, the University of Portsmouth, SLAC National Accelerator Laboratory, Stanford University, the University of Sussex, and Texas A\&M University.
Based in part on observations at Cerro Tololo Inter-American Observatory, National Optical Astronomy Observatory, which is operated by the Association of Universities for Research in Astronomy (AURA) under a cooperative agreement with the National Science Foundation.

\bibliographystyle{JHEP}
\bibliography{references.bib}

\appendix

\section{Galaxy bias function from the Science Verification sample}\label{ap:SV_bias}

\begin{figure}
    \centering
    \includegraphics[width=\linewidth]{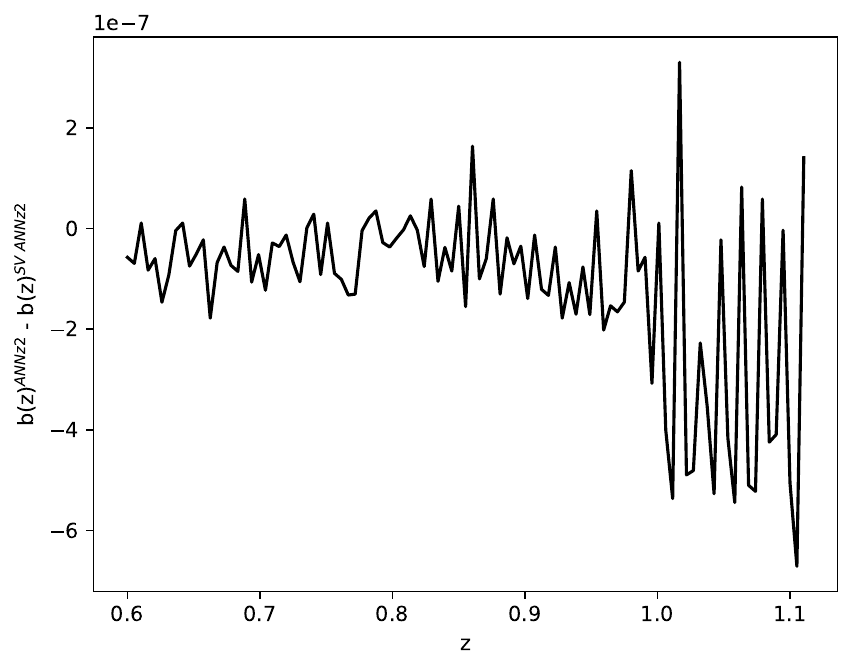}
    \caption{Galaxy bias discrepancy between our data set and the SV data set (also using \texttt{ANNz2}).}
    \label{fig:svbias}
\end{figure}

Due to a significant discrepancy between the Science Verification (SV) data set \cite{salvador2019measuring,crocce2016galaxy}, we conducted a test of equation \ref{eq:biasf} using our \texttt{ANNz2} data and the SV's \texttt{ANNz2} output. The outcomes, illustrated in Figure~\ref{fig:svbias}, reveal a difference on the order of $\mathcal{O}(10e-7)$, which becomes more pronounced at higher redshifts. Consequently, this function is suitable for application to the DES Y3 data set and the Mocks we constructed.

\section{BAO from the parametrized model}\label{ap:sanchez}

We perform Markov Chain Monte Carlo (MCMC) sampling using \texttt{emcee} \cite{foreman2013emcee}. We used flat priors for the following parameters: $5500<B<15000$,  $0<\sigma<15$, $0<C<1$, and $90<s_\perp^{\rm fit}<110$. Because we recomputed the photometric redshifts of the galaxies using a different reference set of spectroscopic redshifts than the DES Collaboration's default, our analysis would, in principle, require mock catalogs tailored to each sample. The goodness-of-fit can be found in table~\ref{tab:chi20}, the best result is \texttt{ANNNz2} Gaussian, closer to unitary, while the worst fit seems to be \texttt{DNF}.

\begin{table}[ht]
    \centering
    \begin{tabular}{c|c}
        Sample & $\chi^2/dof$ \\
        \hline
         \texttt{ANNz2} & 8.15\\
         \texttt{ANNz2} Sp & 5.37\\
         \texttt{ANNz2} Gaussian & 1.21 \\
         \texttt{DNF} & 6.50\\
    \end{tabular}
    \caption{$\chi^2/dof$ for each sample.}
    \label{tab:chi20}
\end{table}

However, the available mocks are designed as statistical realizations of large-scale structure (LSS) and do not account for variations in redshift estimation, they follow equation~\ref{eq:n_spec_des} rather than $z_{p}$. 
Using these mocks to estimate covariance would artificially reduce the error bars on $\xi$, since the mocks do not represent the redshift uncertainties inherent to our analysis.

\begin{figure}
    \centering
    \includegraphics[width=\linewidth]{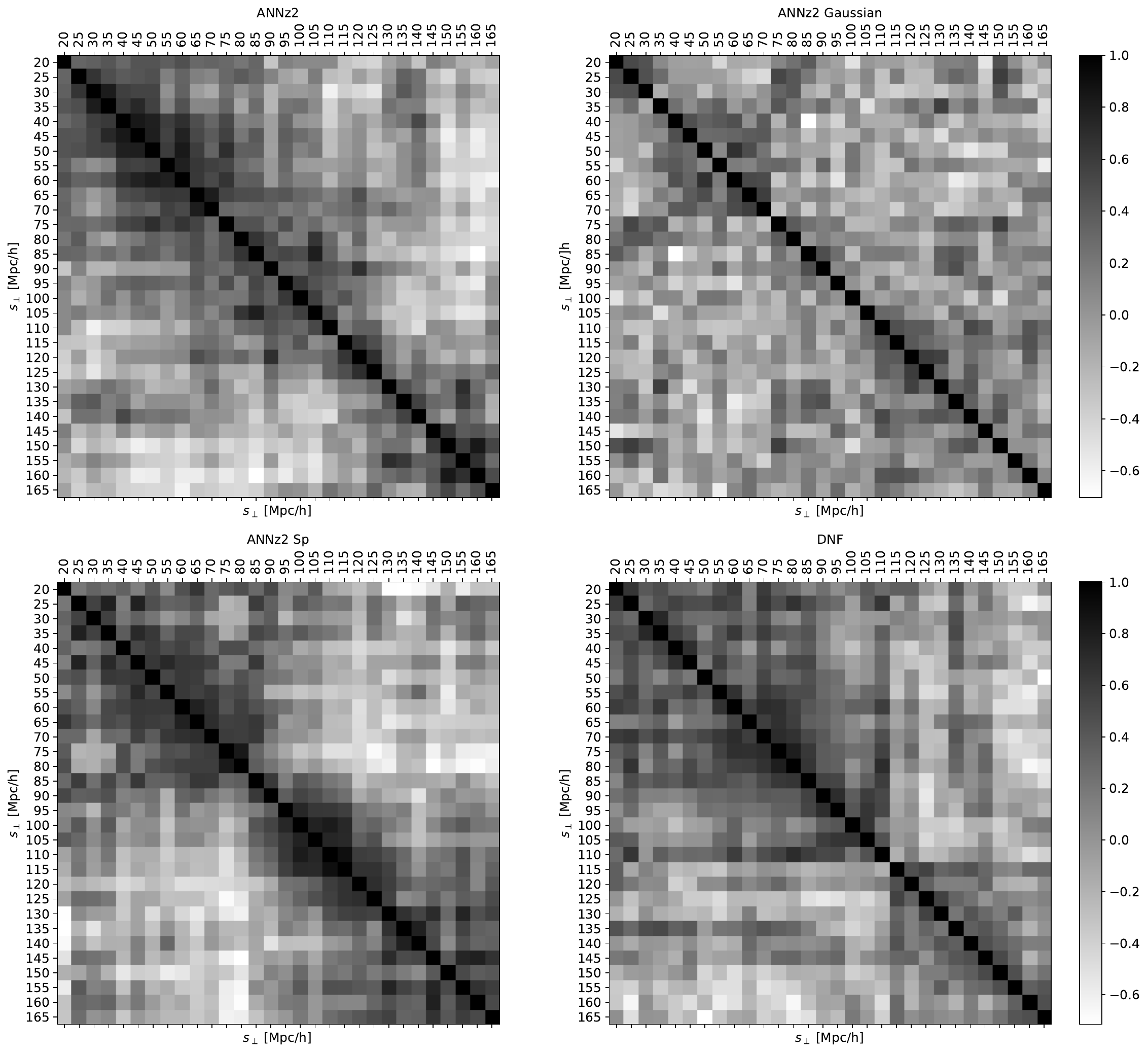}
    \caption{The correlation matrix is associated with the covariance matrices in various configurations: the upper left section represents \texttt{ANNz2}, the upper right pertains to \texttt{ANNz2} Gaussian, the lower left corresponds to \texttt{ANNz2} Sp, and the lower right displays \texttt{DNF}.}
    \label{fig:covariances}
\end{figure}

Our focus is not to test whether DES accurately reproduces the LSS, but rather to quantify how the BAO position depends on redshift estimation methods. 
In order to construct a covariance matrix, we analyse the data by bootstrapping each sample into 50 bins, each with 700,000 galaxies for the full sample, 500,000 for the Sp sample and 200,000 for the Gaussian cuts. The covariance matrices can be found in Figure~\ref{fig:covariances}, it is clear that the \texttt{ANNz2} Gaussian has the least noise covariance matrix.

Crucially, the parameter fits derived from the 100 mocks do incorporate their covariance matrix (estimated from the mock realizations), whereas the data analysis does not. This distinction arises because the mocks are used to infer statistical uncertainties under a fiducial model, while the data analysis focuses on the redshift-dependent BAO signal without assuming the mocks’ covariance is representative of our modified redshift estimates.

We test the BAO detection, by slicing the redshift distribution into five random samples using bootstrap sampling. The results are in Table \ref{tab:chi2}, here we show the $\chi^2$ results from both the best fit of the model in equation \ref{eq:sanchez} but also for the case without BAO ($C = 0$, the null model. \texttt{ANNz2} (with all galaxies) is better than all cases, while \texttt{ANNz} Gaussian is the least precise model due to statistical loss. Compared to the null case, we confirm that the four tested cases present the BAO.

\begin{table}
    \centering
    \begin{tabular}{c|c|c}
        Model&$\chi^2$&Null Model $\chi^2$ \\
        \hline
        \texttt{ANNz2} & 6.41 & 9.60\\
        \texttt{ANNz2} Gaussian & 12.44 & 77.28\\
        \texttt{ANNz2} Sp & 7.09 & 1707.66\\
        \texttt{DNF} & 8.01 & 982.66\\
    \end{tabular}
    \caption{Comparison in goodness of the fit.}
    \label{tab:chi2}
\end{table}

\begin{figure}
    \centering
    \includegraphics[width=\linewidth]{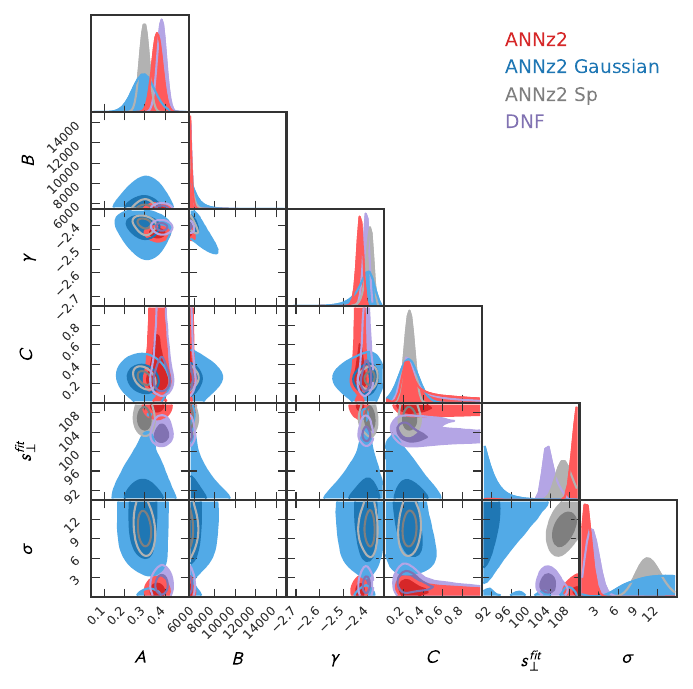}
    \caption{Comparing $\xi(s_\perp$) fitted parameters for \texttt{ANNz2} (\textit{red}: full sample, \textit{blue}: Gaussian, \textit{gray}: Small Peaks) and \texttt{DNF} full sample (\textit{purple}) estimators and after the PDF selection.}
    \label{fig:MCMC}
\end{figure}

In Figure~\ref{fig:MCMC}, we show the resulting constraints for \cite{sanchez2011tracing} parameters. It is clear that , except for the BAO feature, all the samples agree with the correlation function's shape because the parameters A, B, and $\gamma$ are compatible. \texttt{ANNz2} Gaussian's $s_\perp^{fit}$ is significantly different from the other samples, but within 3$\sigma$. 

\section{Error propagation}\label{ap:error}
Here, we give detailed error propagation for $D_M$ and $\alpha$.

The photometric redshift error $\sigma_z$ propagates to the angular diameter distance $D_M$ through the redshift derivative of the distance-redshift relation, where $\sigma_{D_A}$ represents the resulting uncertainty in the distance measurement. This error propagation directly impacts the BAO perpendicular scale measurement $D_M/s^{\rm BAO}_\perp$, with the relative error given by $\sigma_{D_M}/D_M = |dD_M/dz|/D_M \cdot \sigma_z$. The derivative $dD_M/dz$ is cosmology-dependent and typically decreases with increasing redshift, making distance measurements less sensitive to photometric redshift errors at higher redshifts. For typical photo-z errors of $\sigma_z \sim 0.01$--$0.05$, this translates to relative distance errors of approximately 1--10\% across the redshift range most relevant for BAO measurements.

The total uncertainty in the BAO perpendicular scale $D_M/s^{\rm BAO}_\perp$ receives contributions from both the angular diameter distance error $\sigma_{D_M}$ and the sound horizon error $\sigma_{s^{\rm BAO}_\perp}$
\begin{equation}
\sigma_{D_M/r_d} = \sqrt{\left(\frac{\sigma_{D_M}}{s^{\rm BAO}_\perp}\right)^2 + \left(\frac{D_M}{(s^{\rm BAO}_\perp)^2}\sigma_{s^{\rm BAO}_\perp}\right)^2},
\end{equation}
where the first term represents the direct propagation of the distance measurement uncertainty and the second term accounts for the error in the sound horizon scale. This additive error propagation assumes uncorrelated uncertainties between $D_M$ and $s^{\rm BAO}_\perp$, with the sound horizon error term scaling quadratically due to the $1/s^{\rm BAO}_\perp$ dependence in the BAO ratio.

\end{document}